\documentclass[reprint,NumberedRefs]{ARXIVnew}

\usepackage{setspace}
\usepackage{natbib}
\usepackage{graphicx,color,rotating}
\usepackage[latin1]{inputenc}
\usepackage{textcomp}
\usepackage{dcolumn}
\usepackage{bm}     
\usepackage{upgreek}
\usepackage{subdepth}  			
\usepackage{epstopdf}   		
\usepackage[latin1]{inputenc}
\usepackage{amsmath}
\usepackage{amssymb}
\usepackage{amsfonts}
\usepackage{array}
\newcolumntype{L}[1]{>{\raggedright\let\newline\\\arraybackslash\hspace{0pt}}m{#1}}
\newcolumntype{C}[1]{>{\centering\let\newline\\\arraybackslash\hspace{0pt}}m{#1}}
\newcolumntype{R}[1]{>{\raggedleft\let\newline\\\arraybackslash\hspace{0pt}}m{#1}}

\usepackage{gensymb} 

\usepackage{xcolor}						
\definecolor{darkgreen}{rgb}{0.00, 0.50, 0.00}
\definecolor{DARKGREEN}{rgb}{0.00, 0.50, 0.00}
\definecolor{RED}{rgb}{1.00, 0.00, 0.00}
\definecolor{GREEN}{rgb}{0.00, 1.00, 0.00}
\definecolor{BLUE}{rgb}{0.00, 0.00, 1.00}
\definecolor{MAGENTA}{rgb}{1.00, 0.00, 1.00}

\newcommand{\upspace}{\rule{0ex}{3.0ex}}

\newcommand{\mr}[1]{\ensuremath{\mathrm{#1}}}
\newcommand{\myvec}[1]{\bm{#1}}
\newcommand{\ee}{\mathrm{e}}
\newcommand{\ii}{\mathrm{i}}
\newcommand{\dm}{\mathrm{d}}

\newcommand{\iot}{{\ii\omega t}}

\newcommand{\ve}{\varepsilon}
\newcommand{\veO}{\ve_0}

\newcommand{\pp}{\partial}

\newcommand{\nablabf}{\boldsymbol{\nabla}}

\newcommand{\Lapl}{\nabla^2}

\renewcommand{\etal}{{\textit{et~al.}\ }}

\newcommand{\DDD}{\myvec{D}}

\newcommand{\FFFrad}{\myvec{F}^\mathrm{rad}}

\newcommand{\Frad}{F^{\mathrm{rad}}}

\newcommand{\Imat}{\textbf{\textsf{I}}}
\newcommand{\JJJ}{\myvec{J}}

\newcommand{\ks}{k_\mathrm{s}}

\newcommand{\nnn}{\myvec{n}}

\newcommand{\rrr}{\myvec{r}}

\newcommand{\uuu}{\myvec{u}}

\newcommand{\vvv}{\myvec{v}}

\newcommand{\vvvsl}{\vvv^{{}}_\mr{sl}}

\newcommand{\zerovec}{\boldsymbol{0}}

\newcommand{\calF}{\mathcal{F}}

\newcommand{\cfl}{c_\mr{fl}}
\newcommand{\cflsqr}{c^2_\mr{fl}}

\newcommand{\Eac}{E_\mathrm{ac}}

\newcommand{\Eacfl}{E_\mr{ac}^\mr{fl}}

\newcommand{\kapPS}{\kappa_\mathrm{ps}}

\newcommand{\kapfl}{\kappa_\mr{fl}}
\newcommand{\kapcpl}{\kappa_\mr{cpl}}

\newcommand{\Urad}{U^{\mathrm{rad}_{}}}

\newcommand{\epsO}{\epsilon_0}

\newcommand{\etafl}{\eta_\mr{fl}}
\newcommand{\etaflb}{\eta^{{\mr{b}}}_\mr{fl}}
\newcommand{\etacpl}{\eta_\mr{cpl}}
\newcommand{\etacplb}{\eta^{{\mr{b}}}_\mr{cpl}}

\newcommand{\Gamfl}{\Gamma_\mathrm{fl}}

\newcommand{\rhofl}{\rho_\mr{fl}}
\newcommand{\rhocpl}{\rho_\mr{cpl}}
\newcommand{\rhosl}{\rho_\mr{sl}}

\newcommand{\pI}{p_1}

\newcommand{\yO}{y_0}

\newcommand{\rhoPS}{\rho_\mathrm{ps}}

\newcommand{\SICel}{^\circ\!\textrm{C}}
\newcommand{\SICpsm}{\textrm{C}\:\textrm{m$^{-2}$}}

\newcommand{\SIum}{\upmu\textrm{m}}

\newcommand{\SIMHz}{\textrm{MHz}}
\newcommand{\SIkHz}{\textrm{kHz}}
\newcommand{\SIJ}{\textrm{J}}

\newcommand{\SIJpcm}{\textrm{J}\:\textrm{m$^{-3}$}}

\newcommand{\SImuL}{\textrm{\textmu{}L}}

\newcommand{\SIkgm}{\textrm{kg}\:\textrm{m$^{-3}$}}
\newcommand{\SIkgpcm}{\SIkgm}
\newcommand{\SIm}{\textrm{m}}

\newcommand{\SImm}{\textrm{mm}}
\newcommand{\SImum}{\textrm{\textmu{}m}}

\newcommand{\SInm}{\textrm{nm}}

\newcommand{\SIkPa}{\textrm{kPa}}
\newcommand{\SIpTPa}{\textrm{TPa}^{-1}}

\newcommand{\SIGPa}{\textrm{GPa}}
\newcommand{\SIPas}{\textrm{Pa}\:\textrm{s}}
\newcommand{\SImPas}{\textrm{mPa}\:\textrm{s}}

\newcommand{\SIs}{\textrm{s}}
\newcommand{\SIms}{\textrm{ms}}

\newcommand{\SImps}{\SIm\,\SIs^{-1}}

\newcommand{\SIV}{\textrm{V}}

\newcommand{\nn}{\nonumber}
\newcommand{\beq}[1]{\begin{equation} \eqlab{#1}}
\newcommand{\eeq}{\end{equation}}
\newcommand{\bsub}{\begin{subequations}}
\newcommand{\esub}{\end{subequations}}
\def\bal#1\eal{\begin{align}#1\end{align}}
\def\balat#1#2\ealat{\begin{alignat}{#1} #2 \end{alignat}}
\def\bsubal#1 #2\esubal{\bsuba{#1}\begin{align}#2\end{align} \esuba}     
\def\bsubalat#1#2#3\esubalat{\bsuba{#1} \begin{alignat}{#2} #3 \end{alignat} \esuba}
\newcommand{\bsuba}[1]{\bsub \eqlab{#1}}
\newcommand{\esuba}{\esub}

\newcommand{\eqlab}[1]{\label{eq:#1}}
\renewcommand{\eqref}[1]{Eq.~(\ref{eq:#1})}

\newcommand{\figref}[1]{Fig.~\ref{fig:#1}}

\newcommand{\figlab}[1]{\label{fig:#1}}

\newcommand{\secref}[1]{Section~\ref{sec:#1}}

\newcommand{\seclab}[1]{\label{sec:#1}}
\newcommand{\tabref}[1]{Table~\ref{tab:#1}}

\newcommand{\tablab}[1]{\label{tab:#1}}

\newcommand{\sigmabf}{\bm{\sigma}}

\begin{document}

\title[Acoustophoresis in polymer-based  microfluidic devices: modeling and experimental validation]{Acoustophoresis in polymer-based  microfluidic devices: modeling and experimental validation}

\author{Fabian Lickert}
\email{fabianl@dtu.dk}
\affiliation{Department of Physics, Technical University of Denmark,\\
DTU Physics Building 309, DK-2800 Kongens Lyngby, Denmark}

\author{Mathias Ohlin}
\email{mathias.ohlin@acousort.com}
\affiliation{AcouSort AB, Medicon Village, S-223 81 Lund, Sweden.}

\author{Henrik Bruus}
\email{bruus@fysik.dtu.dk}
\affiliation{Department of Physics, Technical University of Denmark,\\
DTU Physics Building 309, DK-2800 Kongens Lyngby, Denmark}

\author{Pelle Ohlsson}
\email{pelle.ohlsson@acousort.com}
\affiliation{AcouSort AB, Medicon Village, S-223 81 Lund, Sweden.}

\date{Published in J.~Acoust.~Soc.~Am.~\textbf{149}, 4281-91 (16 June 2021), doi 10.1121/10.0005113}


\begin{abstract}
A finite-element model is presented for numerical simulation in three dimensions of acoustophoresis of suspended microparticles in a microchannel embedded in a polymer chip and driven by an attached piezoelectric transducer at MHz frequencies. In accordance with the recently introduced principle of whole-system ultrasound resonances, an optimal resonance mode is identified that is related to an acoustic resonance of the combined transducer-chip-channel system and not to the conventional pressure half-wave resonance of the microchannel. The acoustophoretic action in the microchannel is of comparable quality and strength to conventional silicon-glass or pure glass devices. The numerical predictions are validated by acoustic focusing experiments on 5-$\SImum$-diameter polystyrene particles suspended inside a microchannel, which was milled into a PMMA-chip. The system was driven anti-symmetrically by a piezoelectric transducer, driven by a 30-V peak-to-peak AC-voltage in the range from 0.5 to 2.5 MHz, leading to acoustic energy densities of $13~\SIJ/\SIm^3$ and particle focusing times of 6.6~s.
\end{abstract}

\maketitle

\section{Introduction}
\seclab{intro}
Polymer-based microfluidic chips offer a multitude of advantages compared to traditional glass-based devices. A big advantage of polymers is the ease of volume fabrication and the low cost per chip using well-established manufacturing processes such as micro-injection molding or hot embossing. Further processing, such as the creation of channel structures through micro-milling as well as polymer-polymer bonding, can be performed to complete the design. Those processes also bring great flexibility in terms of materials. Thermosoftening plastics such as polycarbonate (PC) or cyclic olefin copolymer (COC), as well as polymethylmethacrylate (PMMA) or polystyrene (PS) are widely seen in the context of microfluidics. The price per polymer chip falls more than an order of magnitude below the typical cost of glass-based devices. This offers a solution to establish acoustophoresis devices also outside academia for use in medical devices. In applications outside the research environment, the need for single use devices rises. Applications such as blood-plasma separation in a point-of-care environment require clean and unused fluidic chips to avoid cross-contamination.  Furthermore, lab-on-a-chip systems are becoming well-established solutions. In order for acoustophoresis to play a role in those systems compatibility with existing polymer-based microfluidic platforms is a requirement.

While polymers are already broadly used in many areas of microfluidics, \citep{Sackmann2014} there have been only a few research groups working with polymers in the field of acoustofluidics. Published work on polymer-based acoustofluidic devices, made of either PMMA or PS, include separation of bacteria and blood cells, \citep{Silva2017, Dow2018}, platelet separation, \citep{Gu2019} purification of lymphocytes,\citep{Lissandrello2018, Dubay2019} as well as particle flow-through separation,\citep{Gonzalez2015, Yang2017}, and focusing.\citep{Moiseyenko2019} A common problem of single channel devices however is the low throughput compared to similar glass or silicon devices. This may be caused by the fact that they typically are designed for an acoustic resonance between the channel walls as is the case for glass or silicon based devices. This assumption is not necessarily true for polymer-based devices where the difference in the acoustic impedance between the chip material and liquid, causing the acoustic reflection, may be much lower. An indication of this is the sometimes surprising optimal operation frequency. \citep{Mueller2013}

Moiseyenko and Bruus recently introduced the principle of whole-system ultrasound resonances (WSUR) \citep{Moiseyenko2019} and contrasted it with the conventional use of bulk acoustic waves (BAW) and surface acoustic waves (SAW) in acoustophoresis devices. According to the WSUR-principle, the optimal conditions for achieving acoustophoresis in polymer devices are obtained by considering the dimensions of the whole system and the corresponding whole-system resonances, instead of attempting to base the acoustophoresis on local standing wave resonances excited locally inside the liquid of the microchannels. We base our analysis of acoustophoresis in polymer chips on the WSUR principle.

In this paper, we present a finite-element model for three dimensional (3D) numerical simulations of polymer-based acoustofluidic devices and validate it experimentally. As a proof of concept, we model acou\-sto\-phoresis of suspended microparticles in a specific microchannel embedded in a PMMA polymer chip and driven by an attached piezoelectric transducer at MHz frequencies. We validate the model experimentally, and use it to explore some of the obstacles for efficient polymer-based acoustophoresis and to design an operational device. Our results show that the usual design rules of conventional glass-based devices do not apply for polymer-based chips. This especially holds true when comparing channel resonances in hard-walled glass devices with the WSUR modes found in polymer-based devices with acoustic impedances close to that of water.

In \secref{device}, we introduce the geometry, the materials, and the design of the polymer-based acoustofluidic device. In \secref{theory}, we present the basic theory, including governing equations and boundary conditions, and its implementation in the numerical 3D finite-element model. We show the resulting fields of the chip at resonance in \secref{sim_results} and define a metric for the efficiency of the acoustophoretic particle focusing as a function of frequency. The experimental setup and the fabricated polymer chip is described in \secref{setup}, and in \secref{experiments} we summarize our experimental findings on the focusing ability of the chip as a function of frequency. Finally, in \secref{conclusion} we conclude with a discussion of the presented results.

\begin{table}[b]
\centering
{\setstretch{0.9}
\caption{\tablab{device_dimensions} The length ($l$), width ($w$), and height ($h$) of the chip (pmma), the channel (ch), the  piezoelectric transducer (pzt), the groove (grv), and the glycerol coupling layer (glc).}
\begin{ruledtabular}
\begin{tabular}{lrclr}
 Symbol            & Value       &        &  Symbol          & Value \\  \hline
 $l_\mr{pmma}$ & $50~\SImm$  &        & $l_\mr{ch}$  & $40~\SImm$ \\
 $w_\mr{pmma}$ & $5~\SImm$   &        & $w_\mr{ch}$  & $375~\SIum$ \\
 $h_\mr{pmma}$ & $1.18~\SImm$& \qquad & $h_\mr{ch}$  & $150~\SIum$ \\
 $l_\mr{pzt}$ & $24~\SImm$  &        & $w_\mr{grv}$ & $300~\SIum$ \\
 $w_\mr{pzt}$ & $8~\SImm$   &        & $h_\mr{grv}$ & $65~\SIum$ \\
 $h_\mr{pzt}$ & $2~\SImm$   &        & $h_\mr{cpl}$ & $20~\SIum$ \\
\end{tabular}
\end{ruledtabular}
}
\end{table}

\section{The Device}
\seclab{device}

The design of our acoustofluidic device is following the design of typical glass-based BAW devices with a long straight channel used for acoustic particle separation.\citep{Barnkob2010, Augustsson2011, Muller2013} As listed in \tabref{device_dimensions}, the channel is rectangular with height $h_\mr{ch} =150~\SImum$ and width $w_\mr{ch} = 375~\SIum$, which in a hard-wall channel would sustain a horizontal pressure half-wave at 2~MHz. In a polymer device governed by WSUR modes, a much different resonance frequency is found. The device consists of a polymer chip made from PMMA, containing a microfluidic channel. Actuation is performed using a piezoelectric lead-zirconate-titanate (PZT) transducer. The PZT transducer and the PMMA chip are coupled through a 20-$\SImum$-thin layer of glycerol ($99\%$ volume-per-volume (v/v) glycerol, $1\%$ v/v water), a well-proven method that allows for long-time operation and access to simple exchange of chip and transducer.\cite{Hammarstrom2010, Lenshof2012, Bode2021} Since the temperature of the device is kept constant in the experiments by using a Peltier-element feedback loop, we neglect thermal effects in the numerical modeling.
\begin{figure}[t!]
 \centering
 \includegraphics[width=\columnwidth]{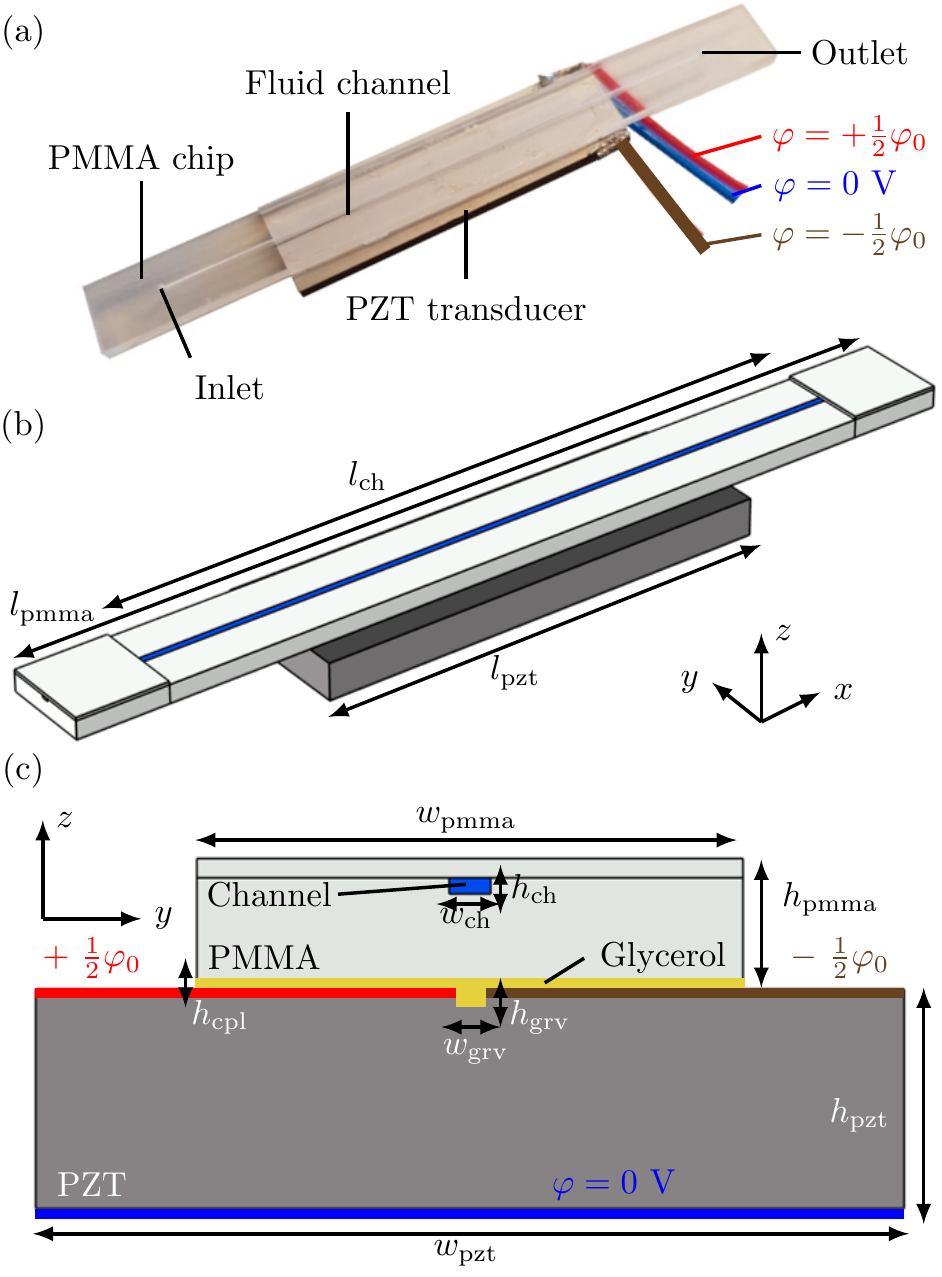}
 \caption[]{\figlab{device_overview}
(a) A photograph of the acoustofluidic device, consisting of PMMA chip with a straight microfluidic channel, a piezoelectric transducer and a coupling layer made from glycerol. (b) Sketch of the 3D model, where the PMMA lid above the channel is removed to show the microchannel (blue) along the $x$-axis. (c) Cross-section of the 3D model in the vertical $yz$ plane. The coupling layer is visualized in yellow.}
 \end{figure}

A sketch of the acoustofluidic device used in the modeling and experiments is shown in  \figref{device_overview} and supplemented by \tabref{device_dimensions}. For simplicity, the shown in- and outlets were omitted in the modeling. To ensure an optimal anti-symmetric motion in the $yz$ plane, the top electrode of the transducer is split in two halves by cutting a small groove using a dicing saw along the $x$-direction and driven by respective AC-voltages with a $180\degree$ phase difference similar to the work reported in the literature.~\cite{Moiseyenko2019, Bode2020, Tahmasebipour2020}

\section{Theory}
\seclab{theory}

\subsection{Governing equations}

In our simulations we follow closely the theory presented by Skov \etal\citep{Skov2019} including the effective boundary layer theory by Bach and Bruus.\citep{Bach2018} We consider a time-harmonic electric potential $\tilde{\varphi}(\rrr,t)$, which excites the piezoelectric transducer and induces a displacement field $\tilde{\uuu}(\rrr,t)$ in the solids as well as an acoustic pressure $\tilde{p}^{{}}_1(\rrr,t)$ in the fluid channel and in the coupling layer,
 \bsubal{harmonic_fields}
 \tilde{\varphi}(\rrr,t) &= \varphi(\rrr)\:\ee^{-\iot},
 \\
 \tilde{\uuu}(\rrr,t) &= \uuu(\rrr)\:\ee^{-\iot},
 \\
 \tilde{p}_1(\rrr,t) &= \pI(\rrr)\:\ee^{-\iot},
 \esubal
with the angular frequency $\omega = 2\pi f$. The time harmonic phase factor $\ee^{-\iot}$ cancels out in the following linear governing equations. From first-order perturbation theory follows that the acoustic pressure $p_\textrm{1,fl}$ in the fluid channel is governed by the Helmholtz equation with damping coefficient $\Gamfl$,
 \beq{EquMotionFluid}
 \Lapl\ p_\textrm{1,fl} = -\frac{\omega^2}{\cflsqr} \big(1+\ii\Gamfl\big)\, p_\textrm{1,fl},
 \text{ with }
 \Gamfl=\Big(\frac{4}{3}\etafl+\etaflb\Big)\,\omega\kapfl,
 \eeq
where $\cfl$ is the speed of sound, $\rhofl$ is the density, $\kapfl = (\rhofl\cflsqr)^{-1}$ is the isentropic compressibility, and $\etafl$ and $\etaflb$ are the dynamic and bulk viscosity of the fluid, respectively. The acoustic velocity $\vvv_\textrm{1,fl}$ of the fluid inside the channel can be expressed as a gradient of the pressure $p_\textrm{1,fl}$ as
 \beq{VelocityPotential}
 \vvv_\textrm{1,fl} = -\ii\:\frac{1-\ii\Gamfl}{\omega\rhofl}\:\nablabf p_\textrm{1,fl}
 \eeq

In the thin glycerol coupling layer, we cannot apply the effective boundary layer theory for the acoustic pressure $p_\textrm{1,cpl}$ and velocity $\vvv_\textrm{1,cpl}$.\citep{Bach2018} So here, we implement the full set of governing equations,
 \bsubal{velocityGlycerol}
 \eqlab{contEquFull}
 \nablabf \cdot \vvv_\textrm{1,cpl} &= \ii\omega\kapcpl p_\textrm{1,cpl},
 \\
 \nablabf \cdot \sigmabf_\textrm{cpl} &= -\ii\omega\rhocpl\:\vvv_\textrm{1,cpl},
 \\
 \sigmabf_\textrm{cpl}  &=\etacpl\big[\nablabf\vvv_\textrm{1,cpl}
 + (\nablabf \vvv_\textrm{1,cpl})^\intercal\big]
 \\ \nn
 &\quad + (\etacplb-\frac{2}{3}\etacpl)(\nablabf\cdot\vvv_\textrm{1,cpl})\Imat
 -p_\textrm{1,cpl}\, \Imat.
 \esubal
Here, $\sigmabf_\textrm{cpl}$ is the viscous stress tensor, $\Imat$ is the identity tensor, $()^\intercal$ is the transpose, $\rhocpl$ is the density, $\kapcpl$ is the isentropic compressibility, and $\etacpl$ and $\etacplb$ are the dynamic and bulk viscosity of the coupling layer, respectively.

The equation of motion for the displacement field $\uuu$ of an elastic solid with density $\rhosl$ is Cauchy's equation
 \beq{CauchyEq}
 -\omega^2\rhosl\:\uuu = \nablabf \cdot \sigmabf_\textrm{sl},
 \eeq
where $\sigmabf_\textrm{sl}$ is the stress tensor. The components $\sigma_{ik}$ of the stress tensor are related by the  stiffness tensor $\bm{C}$ to the strain tensor $\frac{1}{2}(\partial_i u_k + \partial_k u_i)$, which for a linear isotropic elastic material are written in the Voigt notation as
 \beq{StressStrainSolid}
 \resizebox{\columnwidth}{!}{$
 \left( \begin{array}{c}
 \sigma^{{}}_{xx}\! \\[1mm]
 \sigma^{{}}_{yy}\! \\[1mm]
 \sigma^{{}}_{zz}\! \\[1mm]\hline
 \sigma^{{}}_{yz}\! \\[1mm]
 \sigma^{{}}_{xz}\! \\[1mm]
 \sigma^{{}}_{xy}\! \end{array} \right)
  =
 \left( \begin{array}{ccc|ccc}
 C^{{}}_{11} & C^{{}}_{12} & C^{{}}_{12} & 0 & 0 & 0 \\[1mm]
 C^{{}}_{12} & C^{{}}_{11} & C^{{}}_{12} & 0 & 0 & 0 \\[1mm]
 C^{{}}_{12} & C^{{}}_{12} & C^{{}}_{11} & 0 & 0 & 0 \\[1mm] \hline
 0 & 0 & 0  & \!C^{{}}_{44} & 0 & 0 \\[1mm]
 0 & 0 & 0  & 0 & \!C^{{}}_{44} & 0 \\[1mm]
 0 & 0 & 0  & 0 & 0 & \!C^{{}}_{44}
 \end{array} \right) \;
  \left( \begin{array}{c}
 \pp^{{}}_x u^{{}}_x \\[1mm]
 \pp^{{}}_y u^{{}}_y \\[1mm]
 \pp^{{}}_z u^{{}}_z \\[1mm] \hline
 \!\pp^{{}}_y u^{{}}_z\!+\!\pp^{{}}_z u^{{}}_y\! \\[1mm]
 \!\pp^{{}}_x u^{{}}_z\!+\!\pp^{{}}_z u^{{}}_x\! \\[1mm]
 \!\pp^{{}}_x u^{{}}_y\!+\!\pp^{{}}_y u^{{}}_x\! \end{array} \right).$}
 \eeq
Due to symmetry, the remaining three components of the stress are obtained by the relation $\sigma_{ik} = \sigma_{ki}$. The components $C_{ik} = C'_{ik} + \ii C''_{ik}$ of the stiffness tensor $\bm{C}$ are complex-valued to describe the weakly attenuated acoustics in the solid.

The electrical potential $\varphi$ inside the PZT transducer, is governed by Gauss's law for a linear, homogeneous dielectric with a zero density of free charges,
 \beq{GaussLaw}
 \nablabf \cdot \DDD  = 0,
 \eeq
where $\DDD$ is the electric displacement field and $\myvec{\varepsilon}$ the dielectric tensor. Furthermore in PZT, the complete linear electromechanical coupling relating the stress and the electric displacement to the strain and the electric field is given by the Voigt notation as,
 \beq{StressStrainPiezo}
 \resizebox{\columnwidth}{!}{$
 \left( \begin{array}{c}
 \sigma^{{}}_{xx} \\  \sigma^{{}}_{yy} \\  \sigma^{{}}_{zz} \\ \hline
 \sigma^{{}}_{yz} \\  \sigma^{{}}_{xz} \\  \sigma^{{}}_{xy} \\ \hline
 D^{{}}_x \\ D^{{}}_y \\ D^{{}}_z \\
 \end{array}  \right)
 =
 \left( \begin{array}{c@{\:}c@{\:}c@{\:}|c@{\:}c@{\:}c@{\:}|c@{\:}c@{\:}c}
 C^{{}}_{11} & C^{{}}_{12} & C^{{}}_{13} & 0 & 0 & 0 & 0 & 0 & -e^{{}}_{31} \\
 C^{{}}_{12} & C^{{}}_{11} & C^{{}}_{13} & 0 & 0 & 0 & 0 & 0 & -e^{{}}_{31} \\
 C^{{}}_{13} & C^{{}}_{13} & C^{{}}_{33} & 0 & 0 & 0 & 0 & 0 & -e^{{}}_{33} \\ \hline
 0 & 0 & 0 & \!C^{{}}_{44} & 0 & 0 & 0 &  -e^{{}}_{15} & 0 \\
 0 & 0 & 0 & 0 & C^{{}}_{44} & 0 & -e^{{}}_{15} & 0 & 0 \\
 0 & 0 & 0 & 0 & 0 & C^{{}}_{66}  & 0 & 0 & 0 \\ \hline
 0 & 0 & 0 & 0 &  e^{{}}_{15} & 0 & \ve^{{}}_{11} &  0 & 0 \\
 0 & 0 & 0 &  e^{{}}_{15} & 0 & 0 & 0 & \ve^{{}}_{11} & 0 \\
 e^{{}}_{31} & e^{{}}_{31} & e^{{}}_{33} & 0 & 0 & 0  & 0 & 0 & \ve^{{}}_{33}\\
 \end{array}  \right) \;
 \left(   \begin{array}{c}
 \pp^{{}}_x u^{{}}_x \\ \pp^{{}}_y u^{{}}_y \\ \pp^{{}}_z u^{{}}_z \\ \hline
 \pp^{{}}_y u^{{}}_z +\!  \pp^{{}}_z u^{{}}_y \\ \pp^{{}}_x u^{{}}_z +\! \pp^{{}}_z u^{{}}_x  \\ \pp^{{}}_x u^{{}}_y +\!  \pp^{{}}_y u^{{}}_x \\ \hline
 -\pp^{{}}_x \varphi \\ -\pp^{{}}_y \varphi \\ -\pp^{{}}_z \varphi \\
 \end{array}   \right).$}
 \eeq
As before, the remaining three components of the stress tensor are given by the symmetry relation $\sigma_{ik} = \sigma_{ki}$.

\subsection{Boundary conditions between liquid, solid, and PZT}

In the following, we state the boundary conditions of the fields on all boundaries and interfaces of the model. On the surfaces facing the surrounding air, we assume zero stress on the PMMA and the PZT as well as zero free surface charge density on the PZT. On the surfaces with electrodes, the PZT has a specified AC-voltage amplitude. On the internal surfaces between PMMA and PZT, the stress and displacement are continuous, and likewise on the fluid-solid interface, but here in the form of the effective  boundary conditions derived by Bach and Bruus.\citep{Bach2018}  These effective boundary conditions include the viscous boundary layer analytically, and thus we avoid resolving these very shallow boundary layers numerically. The effective boundary conditions include the velocity $\vvvsl = -\ii \omega \uuu$ of the solid (sl) and the complex-valued shear-wave number $\ks = (1+\ii)\:\delta_\textrm{fl}^{-1}$ of the fluid (fl), where $\delta_\textrm{fl}=\sqrt{2\etafl/(\rhofl\omega)} \approx 0.5~\SImum$ is the thickness of the boundary layer. In the coupling layer of height $h_\textrm{cpl} = 20~\SImum$, the boundary layer thickness, $\delta_\textrm{cpl} = 12~\SImum$, is nearly the same, so the effective boundary conditions do not apply. We therefore implement the full continuous conditions for stress and velocity at the interface of the solid (sl) and the coupling layer (cpl),
 \bsubalat{bcAll}{2}
 & \text{PZT bot:}
 & \varphi &= 0,
 \\
 \eqlab{bcPZTphi}
 & \text{PZT top:}
 & \varphi &= \pm \tfrac12 \varphi_0,
 \\
 & \text{PZT-air:}
 &\DDD\cdot\nnn &= 0,
 \\
 & \text{sl-air:}
 & \sigmabf_\textrm{sl} \cdot \nnn &= \zerovec,
 \\
 & \text{sl-fl:}
 & \sigmabf_\textrm{sl} \cdot \nnn & = - p_\textrm{1,fl}\: \nnn + \ii\ks\etafl(\vvvsl - \vvv_\textrm{1,fl}\big),
 \\
 & \text{fl-sl:}
 &\vvv_\textrm{1,fl} \cdot \nnn  &=  \vvvsl\cdot\nnn + \frac{\ii}{\ks} \nablabf_\parallel\cdot\big(\vvvsl - \vvv_\textrm{1,fl}\big)_\parallel,
  \\
 & \text{cpl-sl:}
 & \vvv_\textrm{1,cpl} &= \vvvsl
  \\
 & \text{sl-cpl:}
 & \sigmabf_\textrm{sl} \cdot \nnn &= \sigmabf_\textrm{cpl} \cdot \nnn.
 \esubalat
We use the symmetry at the $yz$- and $xz$-plane to reduce the model to quarter size in the domain $x>0$ and $y>0$ allowing for finer meshing and/or faster computations. We apply symmetric boundary conditions at the $yz$-plane $x=0$ and anti-symmetry at the $xz$-plane $y=0$,
\bsubalat{BC_symmetry}{2}
 \text{Symmetry, } & x=0: &&
 \nn \\
 u_x &= 0,\;  &  \sigma_{yx,\mr{sl}} &= \sigma_{zx,\mr{sl}} = 0,
 \\
 v_{x,\mr{cpl}} &= 0,\; & \sigma_{yx,\mr{cpl}} &= \sigma_{zx,\mr{cpl}} = 0,
 \\
 \pp_x p_\mr{1,fl} &= 0,\; & \pp_x \varphi &= 0.
 \\
 \text{Anti-symmetry, } & y=0: &&
 \nn \\
 \sigma_{yy,\mr{sl}} & = 0, & u_x &= u_z = 0,
 \\
 \sigma_{yy,\mr{cpl}} &= 0, &  v_{x,\mr{cpl}} &= v_{z,\mr{cpl}} = 0,
 \\
 p_{1,\mr{fl}} &= 0,\; & \varphi &= 0.
 \esubalat

\subsection{Acoustic energy density and radiation force}

The space- and time-averaged acoustic energy density $\Eacfl$ in a fluid in a specified volume $V_\mr{fl}$ is given as the sum of the time-averaged kinetic and compressional energy,
 \beq{EacflDef}
 \Eacfl =  \frac{1}{V_\mr{fl}} \int_{V_\mr{fl}}
 \bigg[ \frac14 \rhofl \big| \vvv_\mr{1,fl}\big|^2 + \frac14 \kapfl \big|p_\mr{1,fl}\big|^2 \bigg]
 \:\dm V.
 \eeq
The acoustic radiation force $\FFFrad$ acting on particles in the fluid is minus the gradient of the potential $\Urad$, specified for particles with radius $a$, density $\rhoPS$, and compressibility $\kapPS$, suspended in a fluid with density $\rhofl$ and compressibility $\kapfl$,\cite{Settnes2012}
 \bsubal{FradEq}
 \FFFrad &= -\nablabf \Urad,
 \\
 \Urad &=
 \pi a^3\Big(\frac13 f_0\: \kapfl  |p_\mr{1,fl}|^2
 - \frac12 f_1\:\rhofl |\vvv_\mr{1,fl}|^2 \Big), \\
 f_0 &= 1 - \frac{\kapPS}{\kapfl}, \qquad
 f_1 = \frac{2(\rhoPS-\rhofl)}{2\rhoPS + \rhofl},
 \esubal
where, $f_0$ and $f_1$ is the so-called acoustic monopole and dipole scattering coefficient, respectively.

\subsection{Electrical impedance and admittance}

The electrical impedance $Z = \varphi_0/I$ and admittance $Y = I/\varphi_0$ of the device is defined by the potential difference $\varphi_0$ between the two split top electrodes of the PZT, \eqref{bcPZTphi}, and the electrical current $I$ through one of these electrodes. Denoting the surface of the positive split electrode as $\pp\Omega_+$, we use the surface integral of the current density $\JJJ$ to obtain $I = \int_{\pp\Omega_+} \nnn \cdot \JJJ\:\dm a = -\ii\omega\int_{\Omega_+} \nnn \cdot (\DDD +\epsO\nablabf\varphi)\:\dm a$,\citep{Skov2019b}
 \bal
 \eqlab{impedanceSim}
 & Y = \frac{1}{Z} =  \frac{I}{\varphi_0} = \frac1{\varphi_0} \int_{\pp\Omega_{+}}\JJJ\cdot\nnn\:\dm a =
 \\ \nn
 & \frac{-\ii \omega}{\varphi_0} \int_{\pp\Omega_{+}}
 \!\!\big[e_{31}(\pp_xu_x\!+\!\pp_yu_y) + e_{33}\pp_z u_z
 + (\varepsilon_0\!-\!\varepsilon_{33})\pp_z\varphi\big]\dm a.
 \eal

\subsection{Material properties}
\seclab{material_param}

The values of the material parameters are taken from the literature to match the validation experiments we have carried out. We study a suspension of $4.8~\SIum$-diameter polystyrene particles at a temperature of $T = 20~\SICel$. To obtain neutral buoyancy, the liquid in the microchannel is chosen to be water mixed with a volume fraction of 16\% iodixanol. The polymer is PMMA, the transducer is PZT Pz26, and the coupling layer is glycerol. All parameter values used in the simulation are listed in \tabref{material_values}.

\begin{table}[ht!]
\centering
{\setstretch{1.0}
\caption{\tablab{material_values} List of parameters at $20~\SICel$ used in the numerical simulation. Channel fluid (84\% v/v water, 16\% v/v iodixanol), $4.8~\SIum$-diameter polystyrene particles, glycerol solution (99\% v/v glycerol, 1\% v/v water), PMMA, and PZT. For PMMA $C_{12} = C_{11} - 2C_{44}$. For PZT $C_{12}  = C_{11} - 2C_{66}$. $\veO$ is the vacuum permittivity.
}
\begin{ruledtabular}
\begin{tabular}{lcrc}
 Parameter &  Symbol  & Value & Unit
 \\
 \hline
 \multicolumn{4}{l}{\textit{Water-iodixanol mixture} \citep{Muller2014, Karlsen2016}} \upspace \\
 Mass density & $\rhofl$ & $1050$ & $\SIkgpcm$ \\
 Speed of sound & $\cfl$ & $1482.3$ & $\SImps$ \\
 Compressibility & $\kapfl$ & $433.4$ & $\SIpTPa$ \\
 Dynamic viscosity & $\etafl$ & $1.474$ & $\SImPas$ \\
 Bulk viscosity & $\etaflb$ & $1.966$ & $\SImPas$
 \\
 \hline
 \multicolumn{4}{l}{\textit{Polystyrene} \citep{Karlsen2015}} \upspace  \\
 Mass density & $\rhoPS$ & $1050$ & $\SIkgpcm$ \\
 Compressibility & $\kappa_\mr{ps}$ & $238$ & $\SIpTPa$ \\
 Monopole coefficient & $f_0$ & $0.479$ & -- \\
 Dipole coefficient & $f_1$ & $0$ & --
 \\
 \hline
 \multicolumn{4}{l}{\textit{Glycerol} \citep{Slie1966, Negadi2017, Cheng2008}}  \upspace \\
 Mass density & $\rho_\mr{cpl}$ & $1260.4$ & $\SIkgpcm$ \\
 Speed of sound & $c_\mr{cpl}$ & $1922.8$ & $\SImps$ \\
 Compressibility & $\kappa_\mr{cpl}$ & $214.6$ & $\SIpTPa$ \\
 Dynamic viscosity & $\eta_\mr{cpl}$ & $1.137$ & $\SIPas$ \\
 Bulk viscosity & $\eta_\mr{cpl}^\mr{b}$ & $0.790$ & $\SIPas$
 \\
 \hline
 \multicolumn{4}{l}{\textit{PMMA} \citep{Hartmann1972, Christman1972, Sutherland1972, Sutherland1978, Carlson2003, Simon2019, Tran2016}} \upspace  \\
 Mass density & $\rhosl$ & $1186$ & $\SIkgpcm$ \\
 Elastic modulus & $C_{11}$ & $8.934 - \ii 0.100$ & $\SIGPa$ \\
 Elastic modulus & $C_{44}$ & $2.323 - \ii 0.029$ & $\SIGPa$
 \\
 \hline
 \multicolumn{4}{l}{\textit{PZT} \citep{Skov2019, Bode2020, Hahn2015}} \upspace  \\
 Mass density & $\rho_\mr{sl}$ & $7700$ & $\SIkgpcm$ \\
 Elastic modulus & $C_{11}$ & $168 - \ii 3.36$ & $\SIGPa$ \\
 Elastic modulus & $C_{12}$ & $110 - \ii 2.20$ & $\SIGPa$ \\
 Elastic modulus & $C_{13}$ & $99.9 - \ii 2.00$ & $\SIGPa$ \\
 Elastic modulus & $C_{33}$ & $123 - \ii 2.46$ & $\SIGPa$ \\
 Elastic modulus & $C_{44}$ & $30.1 - \ii 0.60$ & $\SIGPa$ \\
 Coupling constant & $e_{15}$ & $9.86 - \ii 0.20$ & $\SICpsm$ \\
 Coupling constant & $e_{31}$ & $-2.8 + \ii 0.06$ & $\SICpsm$ \\
 Coupling constant & $e_{33}$ & $14.7 - \ii 0.29$ & $\SICpsm$ \\
 Electric permittivity & $\ve_{11}$ & $828\epsO \; (1 - \ii 0.02)$ & -- \\
 Electric permittivity & $\ve_{33}$ & $700\epsO \; (1 - \ii 0.02)$ & -- \\
\end{tabular}
\end{ruledtabular}
}
\end{table}

\section{Results of 3D simulations}
\seclab{sim_results}

The simulations were implemented in the finite-element software COMSOL Multiphysics 5.5.\citep{Comsol55} We closely follow the implementation of the numerical model given by Skov \etal~\citep{Skov2019}, where further details on the implementation are given. Using the symmetry conditions presented above, we solved a quarter of the actual 3D geometry and subsequently obtained the full solutions by mirroring the results along the $xz$- and $yz$-plane. The obtained fields are  the potential $\varphi$ in the PZT, the displacement $\uuu$ in all solid materials, and the acoustic pressure fields $p_\mr{1, fl}$ and $p_\mr{1, cpl}$ in the fluid and the coupling layer, respectively. In our time-harmonic simulations we study the frequency range between $0.5~\SIMHz$ and $2.5~\SIMHz$,
around the nominal 1-MHz resonance of the PZT transducer. The simulations were performed on the DTU high-performance cluster computer using shared-memory parallelism with a total of 16 cores and 160 GB of random access memory. The meshing was done with a maximum element size of $h_\mr{fl}^\mr{max} = 70~\SImum $ in the fluid channel, $h_\mr{pzt}^\mr{max}  = 280~\SImum $ in the PZT and $h_\mr{pmma}^\mr{max}  = 200~\SImum $ in the PMMA, and vertically resolving the boundary layer in the coupling layer with 5 elements. The final mesh consists of about 100.000 mesh elements, corresponding to approximately 1.8 million degrees of freedom. The computation time per frequency was about 20 minutes. We have performed a standard mesh-convergence study to ensure that our meshing is adequate.\cite{Muller2012, Skov2019}

\begin{figure}[b]
 \centering
 \includegraphics[width=\columnwidth]{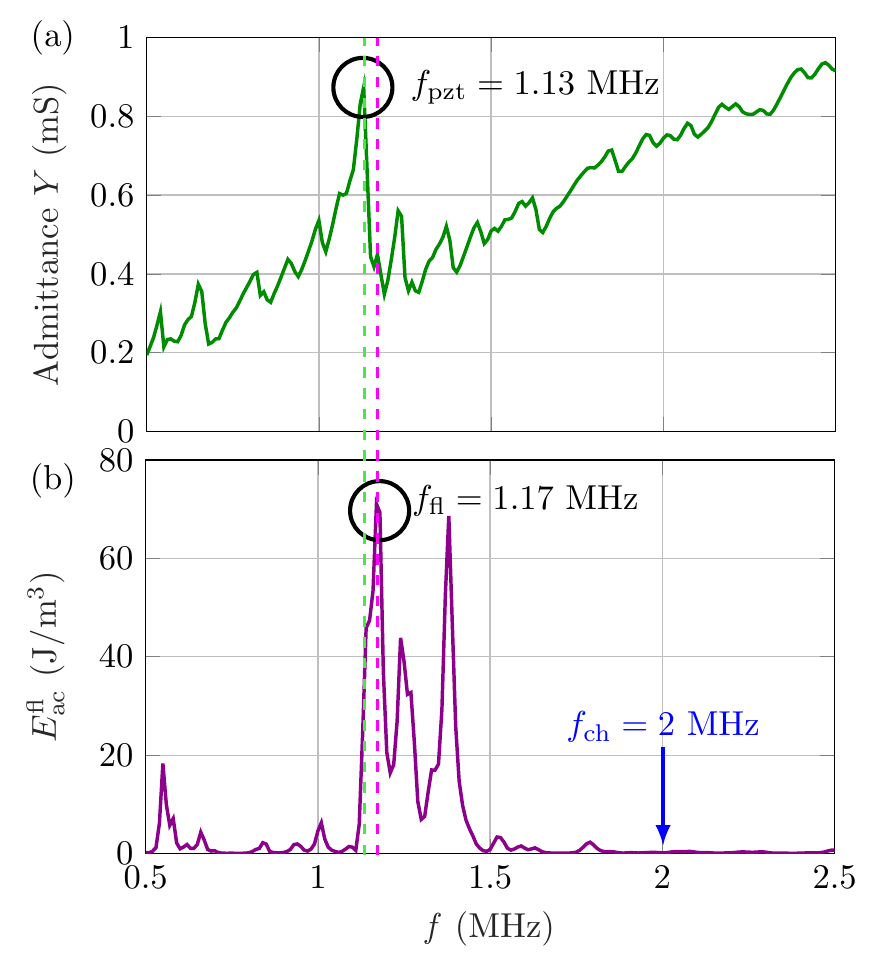}
 \caption[]{\figlab{energy_admittance}
Simulation results in the frequency range 0.5 - 2.5~MHz. (a) The electrical admittance $Y$ of the mounted PZT transducer with a maximum at $f_\mr{pzt} = 1.13~\SIMHz$. (b) The acoustic energy density $\Eacfl$ in the channel with a maximum $\Eacfl = 71~\SIJpcm$ at $f_\mr{fl} = 1.17~\SIMHz$, far below the hard-wall resonance $f_\mr{ch} = 2~\SIMHz$ (blue).}
 \end{figure}

\subsection{Electric admittance and acoustic energy density}
The response of a piezoelectric transducer is usually studied by measuring the electrical impedance $Z$ and finding its characteristic resonance and anti-resonance frequencies. The latter correspond to minima in the electrical impedance spectrum, or maxima in the admittance spectrum  $Y = 1/Z$, and are associated with maxima in the displacement of the transducer.\citep{Bora2015} The simulated electrical admittance spectrum is shown in \figref{energy_admittance}(a). The simulations show a maximum of the admittance at a frequency $f_\mr{pzt} = 1.13~\SIMHz$, close to the $1~\SIMHz$ resonance frequency specified by the manufacturer of the PZT transducer. This fair agreement is obtained despite our use of a split top electrode driven with an anti-symmetric voltage actuation, in contrast to the usual symmetrically driven full-top electrode mode.

\begin{figure*}[t!]
 \centering
 \includegraphics[width=\textwidth]{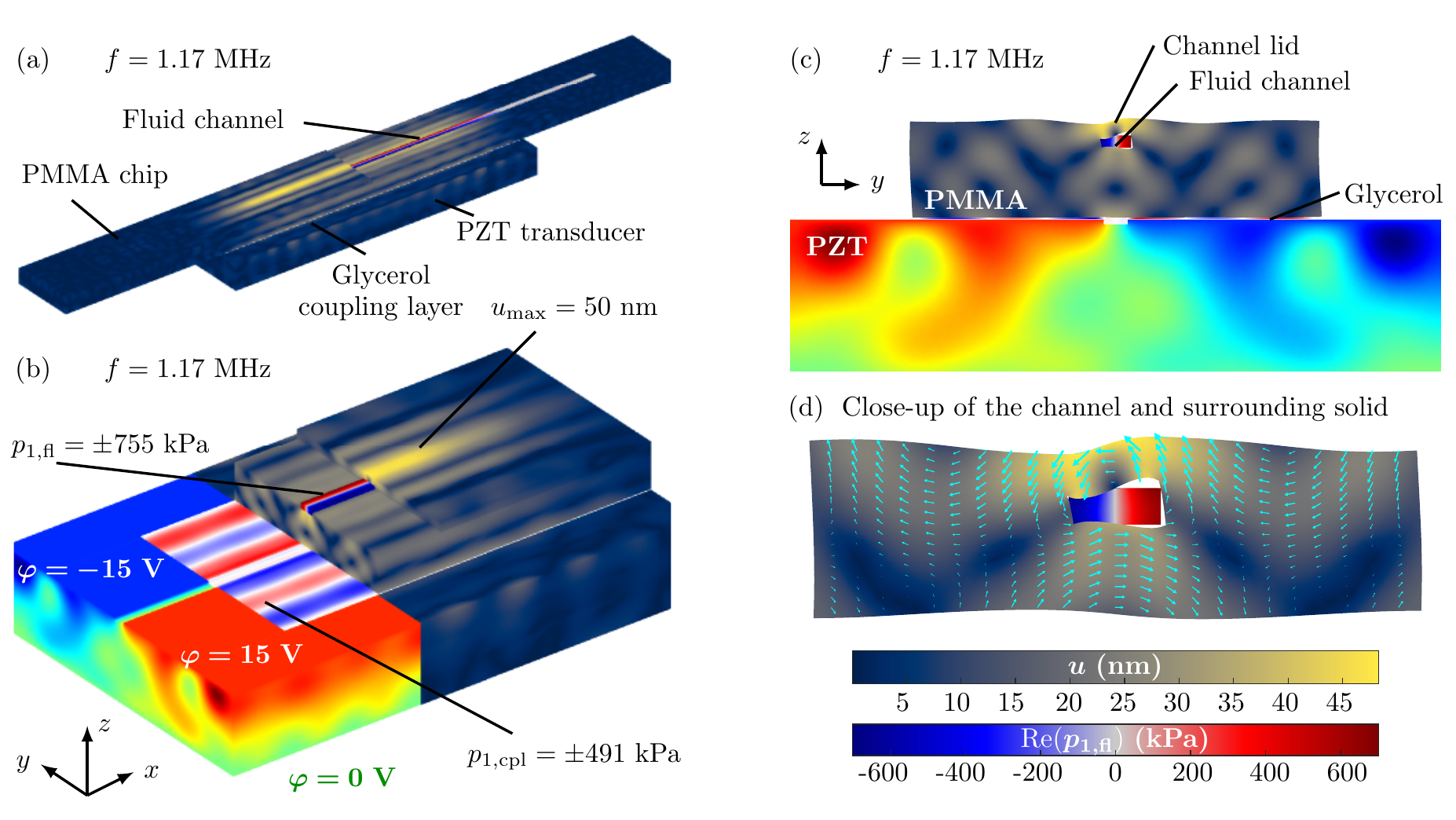}
 \caption[]{\figlab{sim_results_3D}
 Numerical results of the 3D model, evaluated at a frequency of $f = f_\mr{fl} = 1.17~\SIMHz$, and after mirroring the quarter-geometry results in the $yz$- and $xz$-symmetry planes back into the full geometry. (a) Color plot of the displacement magnitude $u$ from $0$ (dark blue) to $50~\SInm$ (yellow), and of the real part of the acoustic pressure $p_\mr{1}$ in the fluid and in the coupling layer from minimum (blue) to maximum (red). (b) Cut-view of the simulated device showing color plots of the fields in the interior parts of the model, including the anti-symmetrically actuated electric potential $\varphi$ in the PZT from $-15~\SIV$ (blue) to $15~\SIV$ (red), and showing the different amplitudes of $p_\mr{1,fl}$ and  $p_\mr{1,cpl}$. (c) Cross section of the device in the $yz$-plane, which emphasizes the motion of the channel lid and the acoustic pressure inside the glycerol coupling layer. (d) Close-up view of the fluid channel and the adjacent lid. The displacement $\uuu$ (cyan vectors) has been scaled with a factor of 1000 to make the lid movement more visible. See the supplementary material for animations of the four views of the resonance mode.\cite{Note1}}
\end{figure*}

In \figref{energy_admittance}(b) is shown the simulated acoustic energy density $\Eacfl$ of that part of the fluid channel, which is located directly above the PZT transducer. We find the maximum value to be $\Eacfl= 71~\SIJpcm$ at $f_\mr{fl} = 1.17~\SIMHz$, which is close to, but $0.04~\SIMHz$ higher than, the resonance frequency $f_\mr{pzt}$ found in the admittance spectrum.

As mentioned in \secref{device}, had the microchannel of width $w_\mr{ch} = 375~\SIum$ had hard walls, it would sustain an acoustic half-wave resonance at $f_\mr{ch} = 2~\SIMHz$. In contrast, the simulations with PMMA walls shows a strong acoustic resonance at $f_\mr{fl} = 1.17~\SIMHz$, much lower than $f_\mr{ch}$, but near the resonance frequency $f_\mr{pzt} = 1.13~\SIMHz$ of the PZT transducer. As the resonance $f_\mr{fl}$ does not match neither $f_\mr{ch}$ nor $f_ \mr{pzt}$,  its is clearly a whole-system resonance.\citep{Moiseyenko2019} This conclusion is supported by a closer inspection of the simulated fields at $f_\mr{fl}$ shown in \figref{sim_results_3D} and in the corresponding videos in the supplementary material.\cite{Note1}

Analyzing the displacement field $\uuu$, we note that the strongest displacement amplitude is obtained in the part of the PMMA located above the PZT, see \figref{sim_results_3D}(d). In particular the highest displacement is found in the region above the fluid channel, which we will refer to as the channel lid in the following. We further note that the acoustic pressure forms a perfect standing anti-symmetric wave (albeit not a half-wave) with a vertical pressure nodal plane along the channel center in the region above the transducer. The amplitude of the pressure in the center of the fluid channel amounts to $p_\mr{1,fl} = 755~\SIkPa$. This pressure amplitude decreases along the $x$-direction, towards both ends of the polymer chip. Finally, we observe a horizontal pressure wave in the glycerol coupling layer with an amplitude of about $p_\mr{1,cpl} =  491~\SIkPa$.

The cross-section of the acoustofluidic device, shown in \figref{sim_results_3D}(c,d), reveals an anti-symmetric motion of the side walls in the horizontal $y$-direction. The channel lid is performing a standing half-wave-like motion, perfectly in phase with the oscillation of the standing pressure wave inside the channel. An analysis of varying geometries of the polymer chip dimensions gave rise to the hypothesis that it is the motion of the side walls, which is driving the channel resonance. In order to obtain a strong resonance it is furthermore important to match the side-wall motion with the motion of the channel lid. Simulations so far have shown ideal results for inward motion of the side wall, coupled with outward motion of the channel lid in one side of the channel. The width $w_\mr{ch}$ of the fluid channel and the thickness of the lid appear to set the frequency of the anti-symmetric standing wave in the lid, and by matching this frequency with that of the anti-symmetric side wall resonance, high acoustic pressure amplitudes and gradients are produced in the channel. This whole-system resonance is governed by the dimensions of the entire geometry of the chip and is difficult to predict analytically.

\subsection{Acoustophoretic focusability}

To predict the acoustic focusing abilities of the device numerically, we compute the fraction of suspended particles focused in the center region of the channel at different focusing times $t$ as a function of frequency. We assume transverse acoustic focusing in the node of a half-wave pressure with the simulated amplitude, neglect acoustic streaming, and consider the case of a neutrally buoyant solution. In this case, the horizontal trajectory $y(y_0, t)$ of a particle at time $t$, starting at position $y_0$, is known analytically. Shifting the coordinate system so that the pressure node is at $y=\frac12 w_\mr{ch}$ and the channel lies at $0 < y < w_\mr{ch}$, we find that\citep{Barnkob2012}
 \bsubal{path}
 y(\yO,t) & = \frac{w_\mr{ch}}{\pi}\arctan\!\Big[\!\tan\!\Big(\pi\frac{\yO}{w_\mr{ch}}\Big) \exp\Big(\frac{t}{t^*}\Big)\Big],
 \\
 t^* &= \frac{3\etafl w_\mr{ch}^2}{4\pi^2\Phi a^2}\frac{1}{\Eacfl},\;
 \text{ with }\; \Phi = \frac13 f_0 + \frac12 f_1,
 \esubal
where $\Phi$ is the acoustic contrast factor and $t^*$ is the characteristic focusing time. Using this expression, we then calculate the fraction of particles that are focused in a band of width $w_\mr{foc}$ around the nodal plane:
\begin{enumerate}
  \item Compute $\Eacfl$ from the numerical simulation and select the focusing band width $w_\mr{foc}$ and time $t_\mr{foc}$.
  \item For a large number $N$ of uniformly distributed initial positions $y_0$ for $0 < y_0 < \frac12 w_\mr{ch}$ compute the final positions $y_\mr{foc} = y(\yO,t_\mr{foc})$ using \eqref{path}.
  \item Count the number $N_\mr{foc}$ of particles inside the focusing band: $y_\mr{foc} > \frac{1}{2}(w_\mr{ch}- w_\mr{foc})$.
\end{enumerate}
The simulated focusability $\calF_\mr{sim}$ is then defined by

\beq{calFsim}
 \calF_\mr{sim} = \frac{N_\mr{foc}}{N}.
\eeq
%

%
\begin{figure}[t!]
 \centering
 \includegraphics[width=\columnwidth]{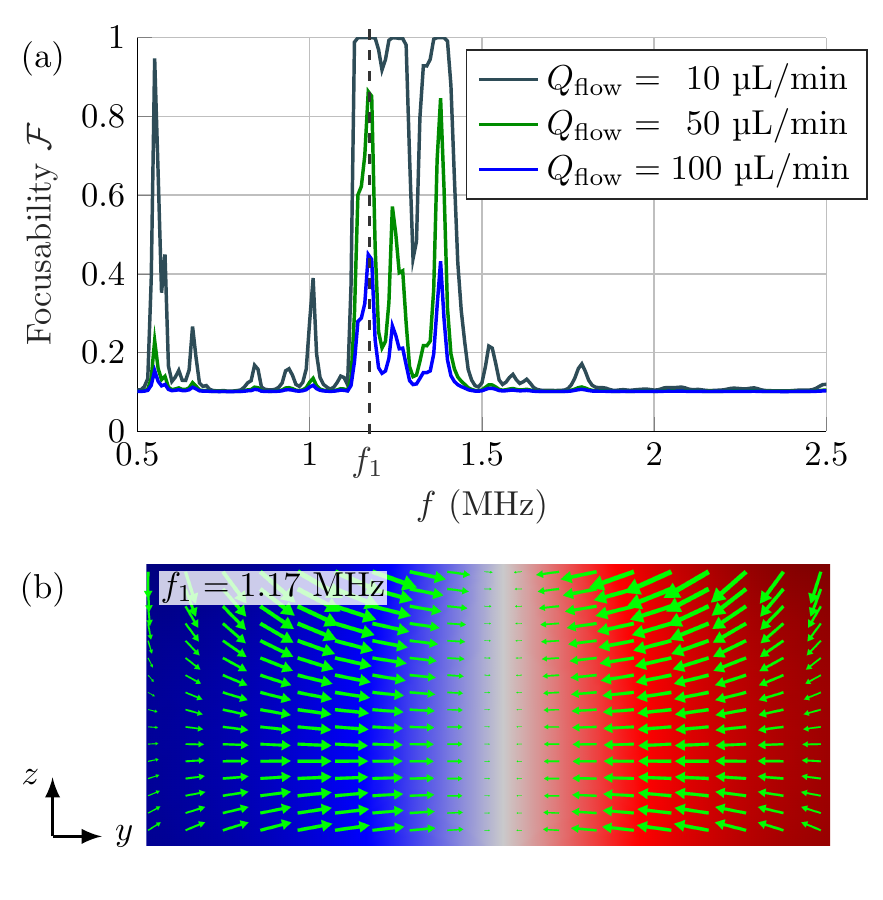}
 \caption[]{\figlab{focusing_2D_channel}
 Simulation results. (a) Plot of the focusability $\calF_\mr{sim}$ versus frequency with a focusing band width $w_\mr{foc} = \frac1{10}w_\mr{ch}$ and for the listed three flow rates $Q_\mr{flow}$. (b) Color plot in the vertical channel cross section of the acoustic pressure $p_\mr{1,fl}$ from $-755$ (blue) to $+755$~kPa (red) and the acoustic radiation force $\FFFrad$ (green vectors) with a magnitude up to $5.5$~pN for suspended 4.8-$\SImum$-diameter polystyrene particles.}
\end{figure}

In our simulations we chose $N = 10^5$ initial positions $y_0$ and a focusing band width $w_\mr{foc} = \frac{1}{10}\,w_\mr{ch}$. We choose the focusing time to be the time it takes a given set flow rate $Q_\mr{flow}$ to sweep half the active volume $V_\mr{fl} = l_\mr{pzt}w_\mr{ch}h_\mr{ch}$ above the PZT transducer,
$t_\mr{foc} = \frac{1}{2}\frac{V_\mr{fl}}{Q_\mr{flow}} = \frac{l_\mr{pzt}w_\mr{ch}h_\mr{ch}}{2Q_\mr{flow}}$, which sets an upper limit to achieve good microparticle focusing in the center of the device for the given geometry. The resulting focusability $\calF_\mr{sim}$ is plotted versus frequency in \figref{focusing_2D_channel}(a) for the three flow rates  $Q_\mr{flow} = 10$, 50, and $100~\SImuL/\mr{min}$, corresponding to the focusing times $t_\mr{foc} = 4.0$, 0.8, and $0.4~\SIs$. The model predicts the best focusing of the device at the frequency $f_1 = 1.17~\SIMHz$, identical to the frequency $f_\mr{fl}$ of \figref{energy_admittance}(b) with the maximum acoustic energy density in the fluid channel. The acoustic pressure $p_\mr{1,fl}$ and the acoustic radiation force $\FFFrad$ at this frequency are shown in \figref{focusing_2D_channel}(b). Clearly, the simulated pressure is an anti-symmetric standing pressure wave, for which the acoustic radiation force points towards the pressure node in the center of the channel causing focusing in the center of the channel of suspended particles. Based on our simulations of the radiation force $\FFFrad = (\Frad_y, \Frad_z)$, we compute the figure of merit,\cite{Moiseyenko2019}  $R = \int_{V_\mr{fl}} -\mr{sign}(y) \Frad_y\:\dm V/\int_{V_\mr{fl}} \big|\Frad_z\big|\:\dm V = 3.9$, which reveals that on average the horizontal focusing force $F_{y}^\mr{rad}$ is about four times larger than vertical force $F_{z}^\mr{rad}$ at the frequency $f_1 = 1.17~\SIMHz$, as can be seen qualitatively from the $\FFFrad$ vectors (green) in \figref{focusing_2D_channel}(b). We therefore concentrate on the focusing in the $y$-direction towards the pressure node in this work.

\section{Experimental setup and results}

\subsection{Setup and procedure}
\seclab{setup}

The first step in the characterization of the acoustofluidic device was the measurement of the electrical admittance. The admittance spectrum $Y(f)$ between the two halves of the split top electrodes is measured using a Digilent Analog Discovery 2 oscilloscope applying the driving voltage to one of the top electrodes, grounding the other top electrode, and leaving the bottom electrode electrically floating. This is equivalent to adding a constant potential $+\frac12 \varphi_0$ to the simulated voltage configuration shown in \figref{device_overview}(c). Observed differences between measured and simulated results for $Y$ might be caused by a temperature sensor which is mounted on one side of the piezoelectric transducer, but not included in the simulations. The piezoelectric transducer is coupled through a thin glycerol layer (99\% v/v glycerol, 1\% v/v water) to the microfluidic polymer chip. The thickness of this coupling layer was measured using a feeler gauge to be approximately $20~\SImum$ thick. For more information about the role of coupling layers, see Refs.~\onlinecite{Hammarstrom2010, Lenshof2012, Bode2021, Bode2021, Hahn2015}.

In the following measurement, a frequency sweep at a fixed voltage amplitude of V = $15~\SIV$ from 0.5 to $2.5~\SIMHz$ was performed to analyze the experimental focusability of a neutrally buoyant suspension of 4.8-$\SIum$-diameter fluorescent polystyrene particles in a water-iodixanol mixture (84\% v/v water, 16\% v/v iodixanol). The solution was pumped through the acoustofluidic device with a flow rate of $10~\SImuL/\mr{min}$ delivered by a syringe pump. Bright-field images were taken in steps of $5~\SIkHz$ with a Hamamatsu Orca Flash 4.0 camera with $50~\SIms$ exposure time. At each frequency the channel was flushed by briefly increasing the flow rate to $1800~\SImuL/\mr{min}$ for $0.1~\SIs$ following by a waiting time of $45~\SIs$ to stabilize the flow at a flow rate of $10~\SImuL/\mr{min}$. Afterwards a series of ten images were taken with the piezoelectric transducer being switched on, and another ten images with the transducer being switched off. The temperature during the experiment was kept constant at $T = 20~\SICel$ using a Peltier element. From the obtained images, an average intensity profile $I_\mr{exp}(y)$ across the channel was calculated at each frequency. The experimental focusability $\calF_\mr{exp}$ was then obtained from the integral of the intensity curve around the channel center divided by the integral across the entire channel, in analogy with $\calF_\mr{sim}$ in \eqref{calFsim},
 \beq{calFexp}
 \calF_\mr{exp}  = \frac{\int_{-\frac12 w_\mr{foc}}^{\frac12 w_\mr{foc}} I_\mr{exp}(y) \, \dm y}{
 \int_{-\frac12 w_\mr{ch}}^{\frac12 w_\mr{ch}} I_\mr{exp}(y) \, \dm y}.
 \eeq
Here we used $w_\mr{foc} = \frac{1}{10}\,w_\mr{ch}$, and thus determined the focusability into a band having the width of 10\% of the channel width $w_\mr{ch}$. In the experiments we observed that at some frequencies there was a small offset from the channel center to the pressure node where the particles got focused. To facilitate the processing of the data in those cases where an intensity offset was observed, we integrated the intensity curve symmetrically around the point $y_\mr{max}$ of maximum intensity, thereby changing the limits of the integral in the numerator to $y_\mr{max} \pm \frac{1}{2}w_\mr{foc}$.

In the final experiment, we measured the acoustic energy density $\Eacfl$ using the same setup as described above: the neutrally buoyant solution, consisting of 84\% (v/v) water, 16\% (v/v) iodixanol and fluorescent 4.8-$\SIum$-diameter polystyrene beads is pumped through the microfluidic polymer chip at $10~\SImuL/\mr{min}$. The same anti-symmetric actuation voltage with an amplitude of $\varphi_0 = \pm15~\SIV$ was used, while the device temperature was kept constant at $T = 20~\SICel$. However, unlike in the previous experiment, only selected frequencies, where some focusing had previously been observed, were studied in this experiment. A series of 600 images was recorded in time steps of $\Delta t = 20~\SIms$, while the fluid flow was stopped. This was done to extract the acoustic energy density $\Eacfl$ around the main resonance frequency from the image series using the light-intensity method presented by Barnkob \etal\citep{Barnkob2012}

\subsection{Experimental results for the electrical admittance, particle focusability, and acoustic energy density}
\seclab{experiments}

We measured the electrical admittance $Y$ as described in the previous section for a 1-MHz PZT transducer, after cutting a groove in the top electrode for anti-symmetric actuation. The admittance was measured while leaving the bottom electrode at a floating potential. The transducer was characterized while coupled to the microfluidic polymer chip. The measured electrical admittance $Y$ and the corresponding acoustic energy density $\Eacfl$ measured at selected frequencies in the range from 0.5 to 2.5~MHz are shown in \figref{adm_energy_plot}.

\begin{figure}[b]
 \centering
 \includegraphics[width=\columnwidth]{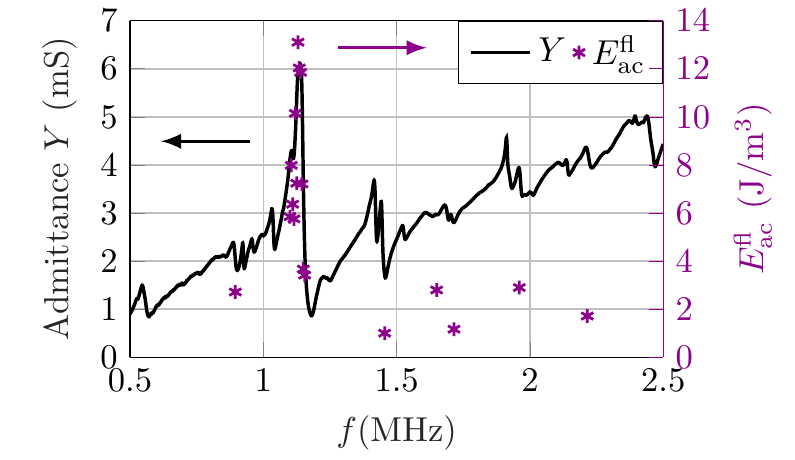}
 \caption[]{\figlab{adm_energy_plot}
The measured admittance spectrum $Y$ (black) from 0.5 to 2.5 MHz using a floating bottom electrode, and the corresponding acoustic energy density $\Eacfl$ (deep purple) obtained by the light intensity method\citep{Barnkob2012} on a series of images recorded under stop-flow condition at selected frequencies showing good particle focusing. The resonance peak in the admittance is located at $f_{Y} = 1.14~\SIMHz$ closely coinciding with the frequency $f_\mr{ac} = 1.13~\SIMHz$, where the energy density attains its maximum value $\Eacfl = 13~\SIJpcm$.}
\end{figure}

The admittance measurement exhibits a strong resonance peak at the frequency $f_Y = 1.14~\SIMHz$. Deviations from the nominal 1-$\SIMHz$-resonance are due to the groove cut into the transducer, the  anti-symmetric actuation, and the load of the chip.  Furthermore, we find that the maximum $\Eacfl = 13~\SIJpcm$ of the acoustic energy density is located close to this maximum of the measured admittance, in good agreement to what is reported in literature for typical glass-based devices.\citep{Bora2015}  This value of $\Eacfl$ corresponds to a focusing time of about $t_\mr{foc} = 6.6~\SIs$ in the channel.

The results of the measurement of the particle focusability  $\calF_\mr{exp}$ during continuous flow operation from 0.5 to 2.5 MHz are shown in \figref{focusability_plot}(a). The frequency with the best focusing is $f_\mr{foc} = 1.13~\SIMHz$, where about $60\%$ of the particles are located within the center $10\%$ of the channel width. Images of the particles inside the channel at this frequency are shown in \figref{focusability_plot}(b) and (c) for the ultrasound switched off and on, respectively. See the supplementary material for a video showing particle focusing.\cite{Note1}

\begin{figure}[t]
 \centering
 \includegraphics[width=\columnwidth]{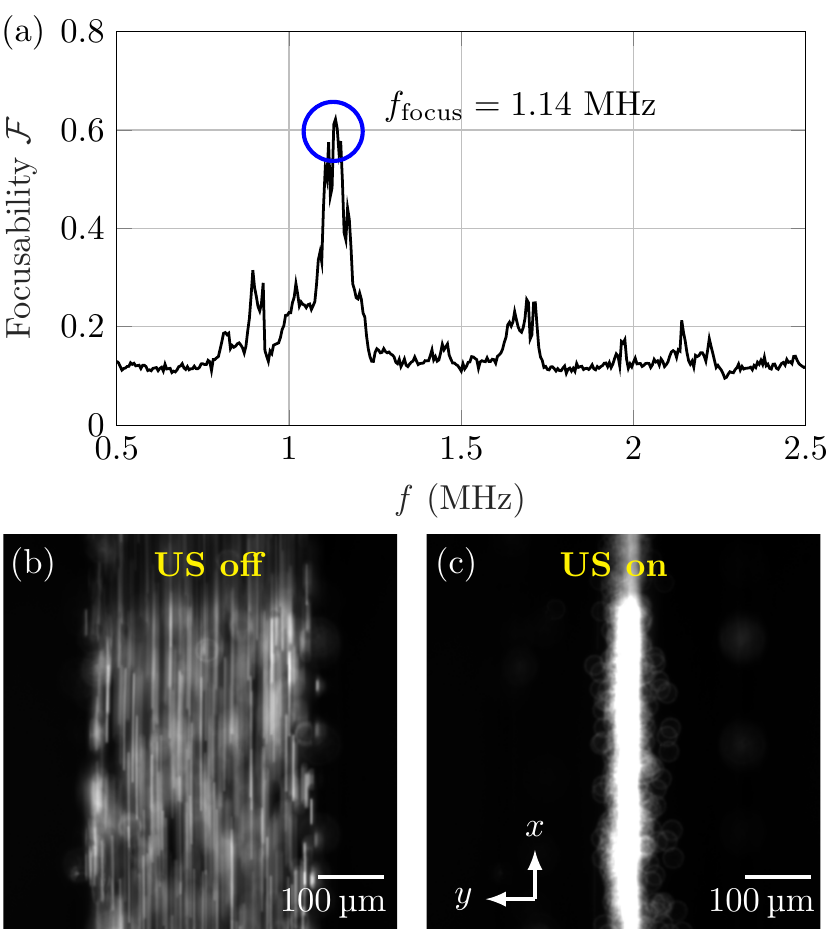}
 \caption[]{\figlab{focusability_plot}
(a) The experimental focusability $\calF_\mr{exp}$ versus frequency, see \eqref{calFexp}, with a maximum 0.6 at $f_\mr{foc} = 1.14~\SIMHz$. (b) Image of the particles in the channel with ultrasound (US) switched off. (c)
Image of the particles in the channel with US switched on at the maximum $f_\mr{foc} = 1.14~\SIMHz$. See the supplementary material for a video showing particle focusing.\cite{Note1}}
\end{figure}

\subsection{Comparison with simulation results}

The simulated values for the three key responses, the admittance $Y$, the focusability $\calF$, as well as the acoustic energy density $\Eac$, agree fairly well with the experimental values. As shown in \figref{compare_exp_sim}(a), the experimental and simulated admittance show the same behavior, and the frequencies of their respective maxima are coinciding within 0.9\%, $f_Y^\mr{exp} = 1.14$~MHz and $f_Y^\mr{sim} = 1.13$~MHz.

\begin{figure}[t]
 \centering
 \includegraphics[width=\columnwidth]{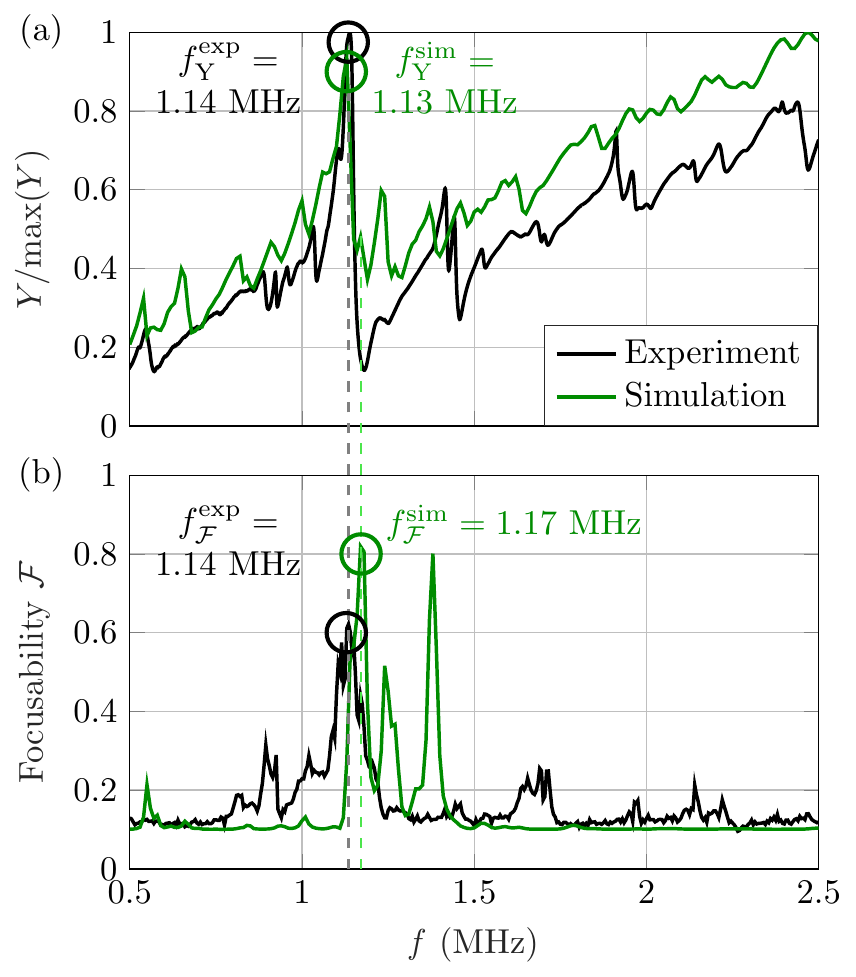}
 \caption[]{\figlab{compare_exp_sim}
 Top-view comparison between measured and simulated responses versus frequency from 0.5 to 2.5~MHz. (a) The measured (black) and simulated (green) electrical admittance $Y$ showing closely coinciding main resonances at $f^\mr{exp}_Y = 1.14$~MHz and $f^\mr{sim}_Y = 1.13$~MHz, respectively. (b) The experimental $\calF_\mr{exp}$ (black) and simulated $\calF_\mr{sim}$ (green) focusability, after calibrating the simulation to match the measured maximum of the acoustic energy density. Both focusabilities $\calF$ show a maximum in the range from 0.6 to 0.8, and a small 2.6\% deviation between the frequencies
 $f^\mr{exp}_\calF = 1.14$~MHz and $f^\mr{sim}_\calF = 1.17$~MHz of the respective maxima.}
 \end{figure}

When comparing the measured maximum value $\Eac^\mr{exp} = 13~\SIJpcm$ of the acoustic energy density with the highest value $\Eac^\mr{sim} = 71~\SIJpcm$ computed in the simulation, we note that the simulation result is about 5.5 times higher than the experimental value. This difference most likely results from neglecting parts of the real system in our idealized simulation, such as tubing, mounting stage, and the inlet and the outlets, all which cause a reduction of the total energy of the real system. We furthermore observe another peak in the simulated acoustic energy density close to the frequency $1.4$~MHz, which has not been observed experimentally. This fact likely stems from a small offset in the $y$-direction between the microfluidic channel and the piezoelectric transducer. This offset could not be implemented in the three-dimensional model as it is breaking the symmetries utilized in the model. Simulations performed in 2D however have shown this peak to decrease drastically with small variations of the chip offset in $y$-direction, while the main peak at $f^\mr{sim}_\calF = 1.17$~MHz stays largely unaffected by this offset.

To compare the simulated and the experimental focusability, we use the standard procedure of calibrating the actuation voltage $\varphi_0$ in the simulation to ensure that $\Eac^\mr{exp} = \Eac^\mr{sim}$.\citep{Muller2013} Using this calibrated actuation voltage, we recalculate the focusability with a flow rate $Q_\mr{flow} = 10~\SImuL/\mr{min}$ according to \eqref{calFsim}, and compare the resulting $\calF_\mr{sim}$ with $\calF_\mr{exp}$ plotted versus frequency in \figref{compare_exp_sim}(b). We observe an upwards frequency shift in the maximum of the simulated focusability curve, here by 2.6\%. Both curves show a similar maximum focusability, namely $\calF_\mr{sim} = 0.82$ for the simulation and $\calF_\mr{exp} = 0.62$ in the experiment. These numbers suggest good focusing of about 60\% to 80\% of the particles. This value can be increased by lowering the flow rate or increasing the voltage amplitude on the transducer. We furthermore note that the highest measured focusability coincides with the global maximum in the measured admittance spectrum, as indicated by the gray-dashed line in \figref{compare_exp_sim}. The simulated maximum in the focusability however relates to a small local maximum in the simulated admittance curve, approximately $40$~kHz above the main admittance resonance $f^\mr{sim}_Y = 1.13$~MHz. This is indicated by the green-dashed line in \figref{compare_exp_sim}. A maximum in the admittance spectrum typically relates to a maximum in the displacement of the piezoelectric transducer, which is driving the whole-system resonance. The small 2.6\% deviation between the frequencies $f^\mr{exp}_\calF$ and $f^\mr{sim}_\calF$ of the focusability maximum likely stems from the idealized assumptions made for \eqref{path}, such as using a perfect horizontal standing half-wave and neglecting the vertical component of the acoustic radiation force. \figref{focusing_2D_channel}(b) and the figure of merit $R = 3.9$ computed in \secref{sim_results}-B show the limitations of this assumption.

\section{Concluding discussion}
\seclab{conclusion}

We have presented a numerical model for 3D simulations of an acoustofluidic polymer device for particle focusing, and we have validated it experimentally. Our 3D simulations predict good acoustic focusing at a frequency of $f_\mr{sim} = 1.17~\SIMHz$, far below the half-wave resonance frequency $f_\mr{ch} = 2~\SIMHz$ corresponding to a rigid hard-wall channel. Furthermore, we observe in our simulations that the resonance in the fluid channel is created through the motion of the side walls in phase with a standing wave motion of the channel lid. It is this whole-system resonance creating the standing pressure half wave, which in turn leads to good focusing at the specified frequency.

In \figref{compare_exp_sim}(a) we find a good qualitative agreement between the simulated and measured electrical admittance spectrum of the device. Quantitatively, only a minor 0.9\% shift in the two spectra was observed. More relevant for applications is the characterization of the ability of the polymer device to focus particles by acoustophoresis. To this end, we have introduced the focusability $\calF$, which can be obtained both by simulation, $\calF_\mr{sim}$ in \eqref{calFsim}, and by experiments, $\calF_\mr{exp}$ in \eqref{calFexp}, thus enabling a good method to compare the two. The focusability $\calF$ is the fraction of the incoming suspended particles, which are focused in the channel center for given focusing times or flow rates, enabling an estimate of the highest achievable flow rates to still maintain reasonable focusing at a selected frequency. Whereas we in \figref{compare_exp_sim}(b) observe a small offset of 2.6\% between the measured and simulated focusability, $\calF_\mr{exp}$ and $\calF_\mr{sim}$ exhibit the same focusing behavior and yields a similar maximum value of $\calF = 0.6 - 0.8$, meaning that 60\% to 80\% of the particles inside the channel are focused in the center 10\% of the channel width.

By studying the electrical admittance, we find both in our simulation and in our experiment that the frequency of the admittance maximum closely coincides with the frequency of the focusability maximum. Both the resonance of the piezoelectric transducer as well as the whole-system resonance are governed by the dimensions of the transducer itself and the whole acoustofluidic device respectively. Whereas the resonance frequency of the piezoelectric transducer is tunable through the height of the transducer, precisely predicting and manipulating the frequency of the WSUR is a more challenging task and requires numerical simulations. Matching this WSUR with the intrinsic resonance frequency of the transducer however would be ideal.

Another approach is to numerically find a design yielding a WSUR at the admittance resonance frequency of the selected piezoelectric transducer. The estimate that can be made based on numerical simulations, however, is only as good as the accuracy of the underlying material parameters. Whereas the mechanical and acoustic properties of glass and silicon are well studied and well reported in literature, it is a challenging task to obtain reliable material parameters for different polymer grades. This especially holds true for data on the transverse speed of sound and attenuation, which are required to compute the complex-valued stiffness coefficient $C_{44}$.

Further studies and measurements beyond the presented proof-of-concept example of the precise properties of the materials in use, will increase the accuracy of our simulation model. Currently, we are working on characterizing various polymers for their applicability as base material in acoustofluidic devices. To fully model the experimental device, fluid connectors and tubing, as well as the clamping of the device in the used measurement setup need to be considered. With our simulation model, however, we obtained a reliable technique to make predictions on the applicability of polymer-based devices for particle focusing applications.

The existing model can be used for further optimizations of the design, in order to yield higher acoustic energy densities and therefore in turn enable flow rates higher than the reported $Q_\mr{flow} = 10~\SImuL/\mr{min}$. Scaling up the flow rate by one or two orders of magnitude seems possible and would make polymer-based acoustofluidic devices competitive with other particle focusing and separation solutions.

\section{Acknowledgments}
This work is part of the Eureka Eurostars-2 E!113461 AcouPlast project funded by Innovation Fund Denmark, grant no.~9046-00127B, and Vinnova, Sweden's Innovation Agency, grant no.~2019-04500.

%
%


\end{document}